\documentclass[preprint,5p]{elsarticle}
\usepackage{amssymb}
\usepackage{amsthm}
\usepackage{bm}
\usepackage[a4paper]{hyperref}
\hypersetup{colorlinks=true,linktoc=all,linkcolor=blue,breaklinks=true,citecolor=blue,urlcolor=blue}
\journal{Journal of Magnetism and Magnetic Materials}

\begin{document}

\begin{frontmatter}

\title{Spin waves in antiferromagnetically coupled bilayers of transition-metal dichalcogenides with Dzialoshinskii-Moriya interaction}

\author[labela]{Wojciech Rudziński}
\ead{wojrudz@amu.edu.pl}

\author[labela,labelb]{Józef Barnaś}
\author[labela]{Anna Dyrdał}

\affiliation[labela]{organization={Faculty of Physics, Adam Mickiewicz University in Poznań},
    addressline={ul. Uniwersytetu Poznańskiego 2},
    city={Poznań},
    postcode={61-614},
    country={Poland}}
\affiliation[labelb]{organization={Institute of Molecular Physics, Polish Academy of Sciences},%Department and Organization
            addressline={ul. M. Smoluchowskiego 17},
            city={Poznań},
            postcode={60-179},
            %state={State Two},
            country={Poland}}

\date{\today}
\begin{abstract}
In this paper we analyze spin waves in bilayers of two-dimensional van der Waals materials, like Vanadium based dichalcogenides, VX$_2$ (X=S, Se, Te) and other materials of similar symmetry. We assume that the materials exhibit Dzialoshinskii- Moriya interaction and in-plane easy-axis magnetic anisotropy due to symmetry breaking induced externally (eg, by strain, gate voltage, proximity effects to an appropriate substrate/oberlayer, etc). The considerations are limited to a collinear spin ground state, stabilized by a sufficiently strong in-plane magnetic anisotropy.
The theoretical analysis is performed within the  general spin wave theory based on the Hollstein-Primakoff-Bogolubov transformation.
\end{abstract}
\end{frontmatter}

\section{Introduction}
In magnetic systems with no inversion symmetry,  the symmetric exchange interactions are accompanied by antisymmetric exchange terms, known as Dzialoshinskii-Moriya interaction (DMI)~\cite{dzial,dzial2,dzial3,mor}. This interaction leads, among others,  to  noncollinear spin textures and nonreciprocal spin wave propagation, [$\omega (\mathbf k)\neq \omega (-\mathbf k)]$~\cite{puszk,seidel,wang,mertig,Ma2014,pirro2017,matan2019,chen2022}.
Recent interest in DMI follows mainly from its role in skyrmion formation in layered metallic structures~\cite{liang,moon}, while the interest in skyrmions is stimulated by possible applications in memory elements~\cite{luo}.

There is currently a broad interest in magnetic van-der-Waals materials, e.g., in Chromium Trihalides (CrI$_3$, CrCl$_3$)~\cite{huang}, chromium ternary tellurides, Cr$_2$X$_2$Te$_6$ (X=Ge, P)~\cite{gong},
transition metal dichalcogenides (TMDs)~\cite{chhow,feng,zhang}, and others.  These materials are naturally 2D layered crystals, that are built of atomic layers with strong in-plain bounds and weak interlayer couplings. Generally, magnetic ordering in van-der-Waals materials depends on the number of monolayers, and  additionally can be easily tuned externally by strain or gating~\cite{Lu2020,Cui2020,Verzhbitskiy2020,Tan2021}. Moreover, they display many interesting phenomena, like  strong magnetoresitance effects~\cite{telford}, spin-to-charge interconversion, topological (electronic and magnon) transport, and others. All this  makes  magnetic van-der-Waals structures very attractive for applications in future spintronics (for a review see, for instance~\cite{Burch2018,Wang22,Li2022}).

Spin waves (magnons) in van dr Waals materials are also of current interest, mainly due to their topological properties~\cite{owerre1,kartsev}, magnon thermal Hall effect~\cite{shen}, and others.  In recent works we have analysed spin wave spectra in bilayers of vanadium-based dichalcogenides VX$_2$ (X=S, Se, Te) in the absence of DMI~\cite{jafariSCRep,rudzPRB} and in monolayers of VX$_2$ with DMI taken into account~\cite{rudzJMMM}. In this paper, we analyze spin-wave spectra of VX$_2$ bilayers, with DMI included. The bilayers are interesting as the corresponding magnon band structure is reminiscent of electronic spectrum in 2D Rashba electron system.
As in Ref~\cite{rudzJMMM}, the considerations in this paper are limited to the case when the collinear ground state is not destroyed by  DMI, which happens when the magnetic anisotropy is sufficiently strong to stabilize the collinear state.
We also assume that the magnetic ground state of an individual monolayer is ferromagnetic, and also assume the easy-plane and in-plane easy-axis anisotropies.
From symmetry arguments follows that DMI and the in-plane easy-axis anisotropy are symmetry-forbidden in a free-standing monolayer of VX$_2$. However, in real systems the symmetry of VX$_2$ can be easily reduced  externally by proximity to a substrate, external gating, strains, etc.  The symmetry also can be  intentionally reduced, as in the so-called   Janus structures, where the monolayer of Vanadium atoms is sandwiched between two monolayers of different chalcogenide atoms~\cite{liang}. All this may lead to nonzero in-plane easy-axis anisotropy and nonzero DMI. Therefore, in this paper we analyse a general model, with both DMI and in-plane easy-axis anisotropy taken into account, and also for spin number $S\ge 1/2$.

To find the spin-wave frequencies we use the  general approach based on the Hollstein-Primakoff-Bogolubov transformation~\cite{owerre1,kamra,brataas}. Accordingly, in section 2  we model the system  under consideration by an appropriate spin Hamiltonian. Then, in section 3 we derive the spin-wave frequencies, while in section 4 we present and discuss numerical results on the spin-wave spectra. Concluding remarks are in section 5.

\section{Model spin Hamiltonian}

We consider a  bilayer of TMDs (e.g. of VX$_2$) in the two stacking geometries, referred to as H (octahedral) and T (trigonal prismatic) ones. The corresponding atomic structure in these two geometries  is shown schematically in Fig.1.
\begin{figure}[hbt!]
\centering
    \includegraphics[width=1.3\columnwidth]{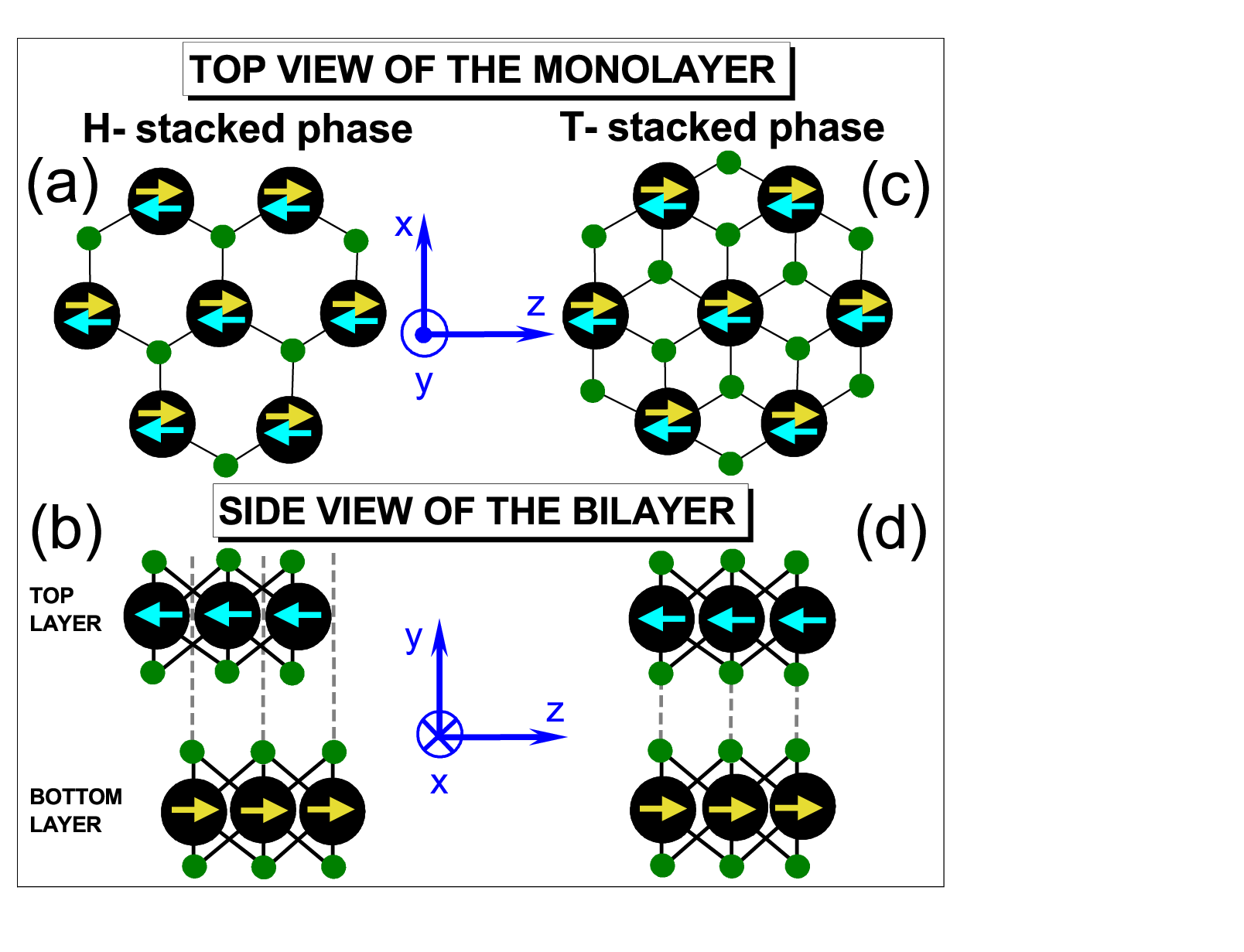}
\caption{(a,c) Atomic structure of a monolayer of TMDs (eg. of VX$_2$) in the H (a) and T (c) phases. The large black dots represent transition metal atoms (e.g., of V), while small green dots represent chalcogen atoms.   The axis $y$ is normal to the atomic plane, while the $x-z$ plane is the magnetic easy plane. The in-plane easy axis is along the $z$ axis.Yellow (blue) arrows represent ground state spin orientation of magnetic atoms along (opposite to) the $z$-axis. (b) Side view of TMDs bilayers in the H (b) and T (d) phases.  In the T phase, the TM atoms of one monolayer are strictly above the TM atoms of the second monolayers, while in the H phase the two monolayers are shifted.}
    \label{Fig:1}
\end{figure}
The bilayer is described by the following spin Hamiltonian:
\begin{equation}
    H=\sum_\alpha H_{\alpha}+H_{\rm int},
\end{equation}
where $\alpha$  = B,T refers to the bottom ($\alpha$ = B) and top ($\alpha$ = T) monolayers, respectively.
Hamiltonian $H_{\alpha}$ represents the individual $\alpha$-th monolayer, while $H_{\rm int}$ describes coupling between the two (B and T) monolayers. Hamiltonian $H_{\alpha}$ can be written  as,
\begin{eqnarray}
    H_{\alpha}=J_1\sum_{\mathbf{r},\bm{\delta}}\mathbf{S}_{\mathbf{r},\alpha}\cdot\mathbf{S }_{\mathbf{r}+\bm{\delta},\alpha}
%     \nonumber \\
     + \frac{D_y}{2}\sum_{\mathbf{r}}\Big(S_{\mathbf{r},\alpha}^y\Big)^2 \nonumber \\
     -\frac{D_z}{2}\sum_{\mathbf{r}}\Big(S_{\mathbf{r},\alpha}^z\Big)^2
%     \nonumber\\
%    -\sum_{\mathbf{r},\mathbf{ \bm \delta}}\mathbf{D}_{\mathbf{r},\mathbf{r}+{\bm {\delta},{\alpha}}}\cdot(\mathbf{S}_{\mathbf{r},{\alpha}}\times\mathbf{S}_{\mathbf{r}+{\bm \delta},{\alpha}})
      +h\sum_{\mathbf{r}}S_{\mathbf{r},\alpha}^z + H_{\rm{DM},\alpha}\;,
\end{eqnarray}
where the first term on the right side stands for the intralayer exchange interactions, the second and third terms describe the easy plane and in-plane easy axis anisotropy, respectively, while the fourth  term is the Zeemann energy in external magnetic field.
The exchange coupling $J_1$ of magnetic atoms  in each  monolayer is assumed ferromagnetic, $J_1<0$. In turn, the easy-plane and in-plane easy-axis anisotropy constants, $D_y$ and $D_z$, are both defined as positive, whereas the external magnetic field $h$ is taken here in energy units.
The sum over $\textbf{r}$ means  here the sum over lattice sites, while that over $\bm{\delta}$ means the sum over nearest neighbours, with $\bm{\delta}$ standing for the vectors connecting a given site to its in-plane nearest neighbours (NNs) (for simplicity, we limit considerations here to the exchange coupling between the nearest neighbours).

The last term in Eq.~(2) describes the intra-layer Dzialo-shinskii-Moriya interaction,
\begin{equation}
  H_{\rm{DM},\alpha}=-\sum_{\mathbf{r},\mathbf{\bm \delta}}\mathbf{D}_{\mathbf{r},\mathbf{r}+{\bm {\delta},{\alpha}}}\cdot(\mathbf{S}_{\mathbf{r},{\alpha}}\times\mathbf{S}_{\mathbf{r}+{\bm \delta},{\alpha}}),
\end{equation}
where the corresponding Dzialoshinskii-Moriya vectors $\mathbf{D}_{\mathbf{r},\mathbf{r}+{\bm \delta},{\alpha}}$ have generally two components in the materials under considerations~\cite{liang},
\begin{equation}
   \mathbf{D}_{\mathbf{r},\mathbf{r+{\bm \delta}},{\alpha}}=d_{\parallel}({\mathbf{\hat{u}}_{\mathbf{r},\mathbf{r}+\bm{\delta},{\alpha}}\times}\mathbf{\hat{y}})+d_{\perp}\mathbf{\hat{y}},
\end{equation}
with
\begin{eqnarray}
    && \bm{\hat{u}}_{\mathbf{r},\mathbf{r}+\bm{\delta}_{1,2},{\alpha}}\times\bm{\hat{y}}={\mp}\bigg(\frac{1}{2}\mathbf{\hat{x}}-\frac{\sqrt{3}}{2}\mathbf{\hat{z}}\bigg),
    \nonumber\\
    & &
    \bm{\hat{u}}_{\mathbf{r},\mathbf{r}+\bm{\delta}_{3,4},{\alpha}}\times\bm{\hat{y}}={\mp}\mathbf{\hat{x}},
    \nonumber\\
    & &
     \bm{\hat{u}}_{\mathbf{r},\mathbf{r}+\bm{\delta}_{5,6},{\alpha}}\times\bm{\hat{y}}={\mp}\bigg(\frac{1}{2}\mathbf{\hat{x}}+\frac{\sqrt{3}}{2}\mathbf{\hat{z}}\bigg).
\end{eqnarray}
It has been  shown in Ref.~\cite{liang}, that the second term in Eq.(4) does not play an important role, and therefore it will be omitted in the following. As already mentioned above,  the ideal monolayer possesses  the inversion center, so the relevant DMI vanishes. It may be induced externally by strain or gate voltage (normal electric field). The DMI appears also inherently in Janus structures, where the magnetic plane  is sandwiched between two monolayers of different atoms~\cite{liang} (which breaks the inversion symmetry).
The DMI Hamiltonian can be written in the  form:
\begin{equation}
   H_{DM}^{\alpha}=-\sum_{\mathbf{r},\mathbf{\bm \delta}}h^{\parallel}_{\mathbf{r},\mathbf{r}+\mathbf{\bm\delta},{\alpha}},
\end{equation}

with
\begin{eqnarray}
   h^{\parallel}_{\mathbf{r},\mathbf{r}+\mathbf{\bm \delta}_{1,2},{\alpha}}=
   \mp\frac{1}{2}d_\parallel\big(S^y_\mathbf{r}S^z_{\mathbf{r}+\mathbf{\bm \delta}_{1,2},{\alpha}}-S^z_\mathbf{r}S^y_{\mathbf{r}+\mathbf{\bm \delta}_{1,2},{\alpha}}\big)
\nonumber\\
    \pm\frac{\sqrt{3}}{2}d_{\parallel}\big(S^x_\mathbf{r}S^y_{\mathbf{r}+\mathbf{\bm \delta}_{1,2},{\alpha}}-S^y_\mathbf{r}S^x_{\mathbf{r}+\mathbf{\bm \delta}_{1,2},{\alpha}}\big),\hspace{0.3cm}
\end{eqnarray}
\begin{eqnarray}
   h^{\parallel}_{\mathbf{r},\mathbf{r}+\mathbf{\bm \delta}_{3,4},{\alpha}}=
   {\mp}d_\parallel\big(S^y_\mathbf{r}S^z_{\mathbf{r}+\mathbf{\bm \delta}_{3,4},{\alpha}}-S^z_\mathbf{r}S^y_{\mathbf{r}+\mathbf{\bm \delta}_{3,4},{\alpha}}\big),
\end{eqnarray}
\begin{eqnarray}
   h^{\parallel}_{\mathbf{r},\mathbf{r}+\mathbf{\bm \delta}_{5,6},{\alpha}}=
   \mp\frac{1}{2}d_\parallel\big(S^y_\mathbf{r}S^z_{\mathbf{r}+\mathbf{\bm \delta}_{5,6},{\alpha}}-S^z_\mathbf{r}S^y_{\mathbf{r}+\mathbf{\bm \delta}_{5,6},{\alpha}}\big)
\nonumber\\
    \mp\frac{\sqrt{3}}{2}d_{\parallel}\big(S^x_\mathbf{r}S^y_{\mathbf{r}+\mathbf{\bm \delta}_{5,6},{\alpha}}-S^y_\mathbf{r}S^x_{\mathbf{r}+\mathbf{\bm \delta}_{5,6},{\alpha}}\big).\hspace{0.3cm}
\end{eqnarray}

Finally, the last term in Eq.(1) represents  the antiferromagnetic exchange coupling between the monolayers,
\begin{equation}
    H_{int}=2J_2\sum_{\mathbf{r},\bm{\delta}}\textbf{S}_{\mathbf{r},T}\cdot\textbf{S}_{\mathbf{r}+\bm{\delta},B},
\end{equation}
with $J_2>0$.  In the above formula, the summation is over lattice sites $\textbf{r}$ in a single monolayer only (therefore, there is a factor of 2 on the right side). Apart from this, $\bm{\delta}$ is here the vector connecting inter-layer NNs.

\section{Spin wave excitations}

We consider  the antiferromagnetic  phase, which appears below the transition field to the spin-flop phase. Therefore, we assume $h=0$. The spin moments of the bottom layer are along $z$ axis, while of the top layer are along the $-z$ axis, see Fig.1.  To find the spin wave frequency we apply the standard procedure based on the Holstein-Primakoff transformation followed by the Fourier transformation and finally Bogoliubov transformation. For more details see Appendix B and also Refs~\cite{owerre1,kamra,brataas}. As a result we obtain frequencies of the two spin wave modes given by the formula,
\begin{eqnarray}
    \omega_{\mathbf{k},\mu} & = & \frac{1}{\sqrt{2}}\bigg{\{}(A_{\mathbf{k}}^+)^2+(A_{\mathbf{k}}^-)^2-2|B_{\mathbf{k}}|^2-8C^2
    \nonumber\\
    & & {\pm} \bigg{\{}(A_{\mathbf{k}}^+-A_{\mathbf{k}}^-)^2\Big[(A_{\mathbf{k}}^++A_{\mathbf{k}}^-)^2-4|B_{\mathbf{k}}|^2\Big]
    \nonumber\\
    & &
    +64C^2|B_{\mathbf{k}}|^2\bigg{\}}^{\frac{1}{2}}\bigg{\}}^{\frac{1}{2}},
\end{eqnarray}
where
\begin{eqnarray}
    A_{\bf{k}}^{\pm}=S\bigg[2J_1\big(\gamma_\mathbf{k}-6\big)+2{\xi}J_2+\frac{D_y}{2}+D_z\nonumber\\
    \pm 4\sqrt{3}d_{||}\gamma_{\mathbf{k}}^{DM}\bigg],
    \\
     B_{\mathbf{k}}=2{\eta}^*_{\mathbf{k}}J_2S, \hspace{1cm} {\rm and} \hspace{1cm}
             C=-\frac{D_yS}{4}.
\end{eqnarray}
Here, the  $+(-)$ sign in $A_{\mathbf{k}}^{\pm}$ corresponds to the bottom (top) layer, while the structure factors $\gamma_\mathbf{k}$, $\xi$,  $\gamma_{\mathbf{k}}^{DM}$ and $\eta_\mathbf{k}$ are defined  as
\begin{equation}
    \gamma_\mathbf{k}=2\bigg(\cos (k_za)+2\cos (\frac{\sqrt{3}}{2}k_xa) \cos (\frac{1}{2}k_za)\bigg),
\end{equation}
\begin{equation}
    \xi = \left\{ \begin{array}{ll}
1 & \textrm{(for T stacking)}\\
3 & \textrm{(for H stacking)},
\end{array} \right.
\end{equation}
\begin{equation}
   \gamma_{\mathbf{k}}^{DM}=\sin\bigg(\frac{\sqrt{3}}{2}k_xa\bigg)\cos\bigg(\frac{1}{2}k_za\bigg),
\end{equation}
\begin{equation}
    \eta_{\mathbf{k}} = \left\{ \begin{array}{ll}
1 & \textrm{(for T phase)}\\
e^{i\frac{k_xa}{\sqrt{3}}}+2e^{-i\frac{k_xa}{2\sqrt{3}}}\cos ({\frac{1}{2}k_za)} & \textrm{(for H phase).}
\end{array} \right.
\end{equation}

Note, the above dispersion relation holds for the collinear ground states.
To find the condition when the above relation is applicable one needs to determine  the magnitude of the anisotropy constant $D_z$ that stabilizes the collinear ferromagnetic ground state.
Assuming $k_z=0$, one can determine the minimum magnon energy from the  condition $\frac{d\omega_{\mathbf{k},-}}{dk_x}=0$.
From this follows that for a given $d_\parallel$, the minimum
in-plane anisotropy $D_z$,
that stabilizes the collinear ferromagnetic ground state, may be evaluated from the condition $\omega_{\bm{k},-}=0$.

\section{Numerical results}

As we already mentioned above, an important property of van der Waals  materials is tunability of some of their parameters due to proximity effects, and by external strain and electric filed. This especially applies to the constants of  Dzialoshinskii-Moryia interaction and of  magnetic anisotropies. Therefore, to emphasize   the DMI-induced
features of the spin wave spectrum,  we assumed in numerical calculations reasonable nonzero values of these parameters, even if they vanish in free-standing bilayers of V-based TMDs.
\begin{figure}[hbt!]
\centering
    \includegraphics[width=0.9\columnwidth]{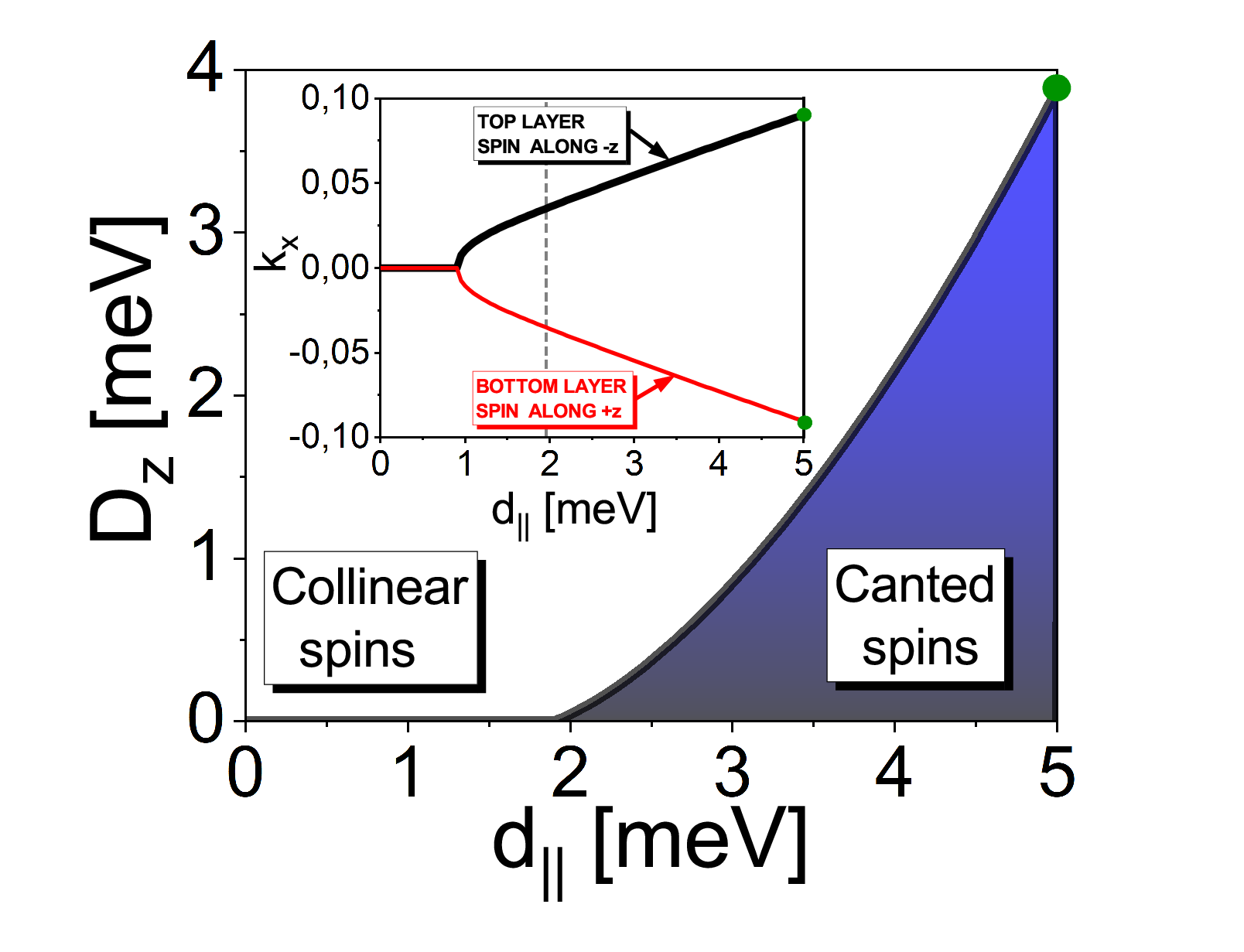}\\
\caption{The phase diagram in the space ($D_z,d_{\parallel}$). The boundary between the shaded area (canted spin states) and white area above (collinear state) defines the minimum value of $D_z$ that stabilizes the collinear state for a given value of DMI parameter $d_\parallel$ (and other parameters as described in the text). The green points correspond to $d_\parallel$ and $D_z$ used in the   numerical calculations. Inset: Black and red lines present wavevectors at which the spin wave energy reaches minimum. The vertical gray dashed line indicates the $d_{||}$ at which the canted-spin phase begins when $D_z=0$.}
    \label{Fig:2}
\end{figure}
\begin{figure*}[hbt!]
\centering
    \includegraphics[width=2.0\columnwidth]{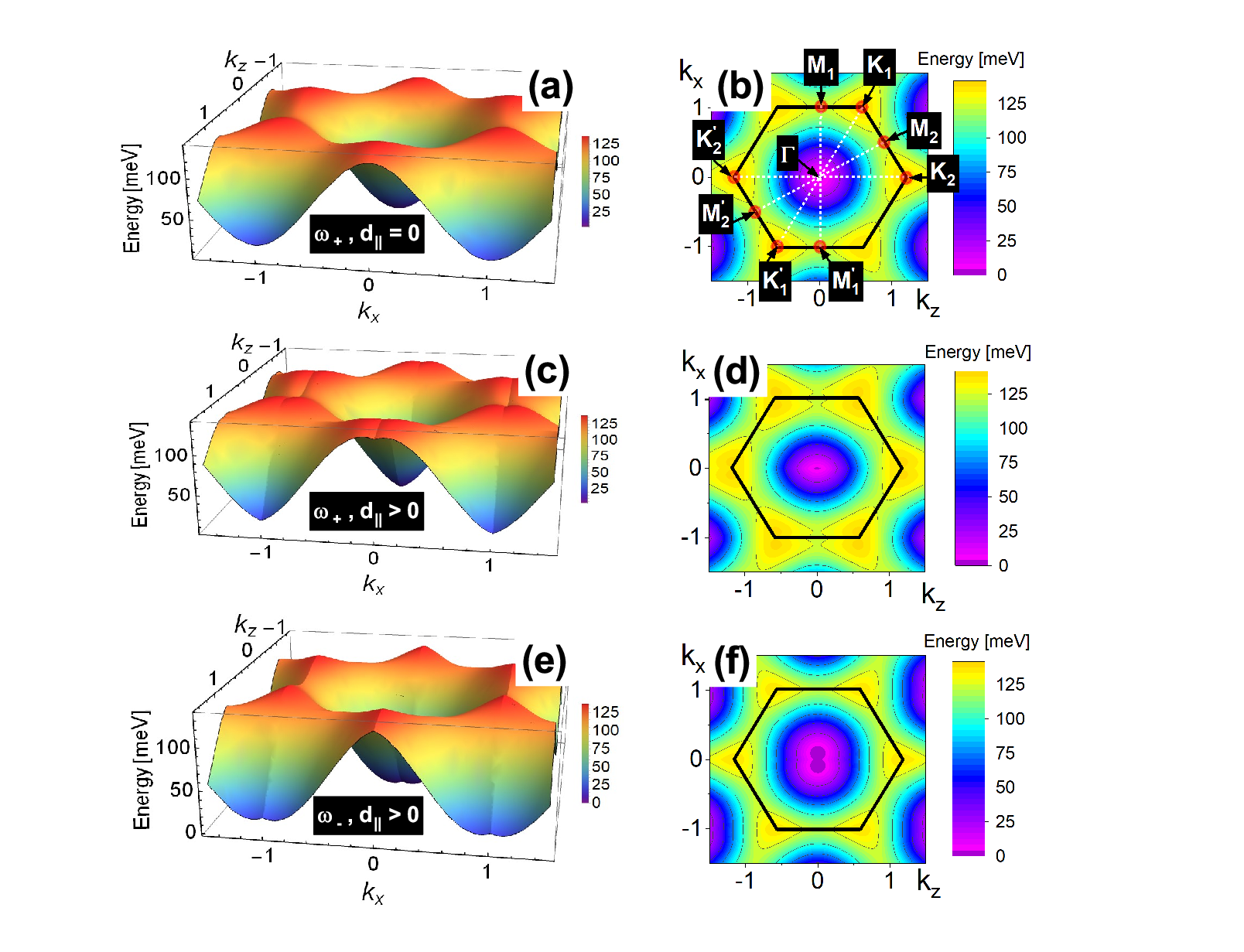}
\caption{Spin wave energy in the whole Brillouin zone for vanishing DMI parameter (a) and for the upper (c) and lower (e) modes in the case of $d_{\parallel}$=5 meV.
The corresponding projections on the $(k_z,k_x)$ plane are shown in the right column, Figs.\ref{Fig:3}(b,d,f).
Other parameters as described in the main text. The white dotted lines in (b) correspond to the crossections, for which the dispersion curves are presented below. In turn, M$_1$, M$_2$, K$_1$ and M$_2$, as well as the corresponding primed symbols, are the relevant points in the Brillouin zone. }
    \label{Fig:3}
\end{figure*}

Accordingly, in the following we assume  $a$=3.59006 ${\AA}$, $J$=-16.52 meV, $D_y$=1.93 meV ~\cite{jafariSCRep},  and $h$=0. To have the collinear ground state, the anisotropy constant $D_z$ will be adjusted for each value of $d_{\parallel}$, to obey the relevant condition.
The relation between  $d_{\parallel}$ and $D_z$ is shown in Fig.2,
where the shaded area corresponds to the canted ground state configuration, while the collinear ground state appears for larger values of $D_z$.  The parameters used in the numerical calculations correspond to the green large points in this diagram.

For the following numerical calculations we assumed $d_{\parallel}$=5 meV and the anisotropy constant $D_z$=4.46 meV (see the green points in Fig.2).
In the left column of Fig.\ref{Fig:3} we show the energy of spin waves in the whole two-dimensional Brillouin zone for zero DMI, Fig.\ref{Fig:3}(a), and for $d_{\parallel}$=5 meV in Fig.\ref{Fig:3}(c) (upper mode) and in Fig.\ref{Fig:3}(e) (lower mode).
The corresponding projections on the $(k_z,k_x)$ plane are shown in the right column, Figs.\ref{Fig:3}(b,d,f).

\begin{figure}[hbt!]
\centering
 \includegraphics[width=1.0\columnwidth]{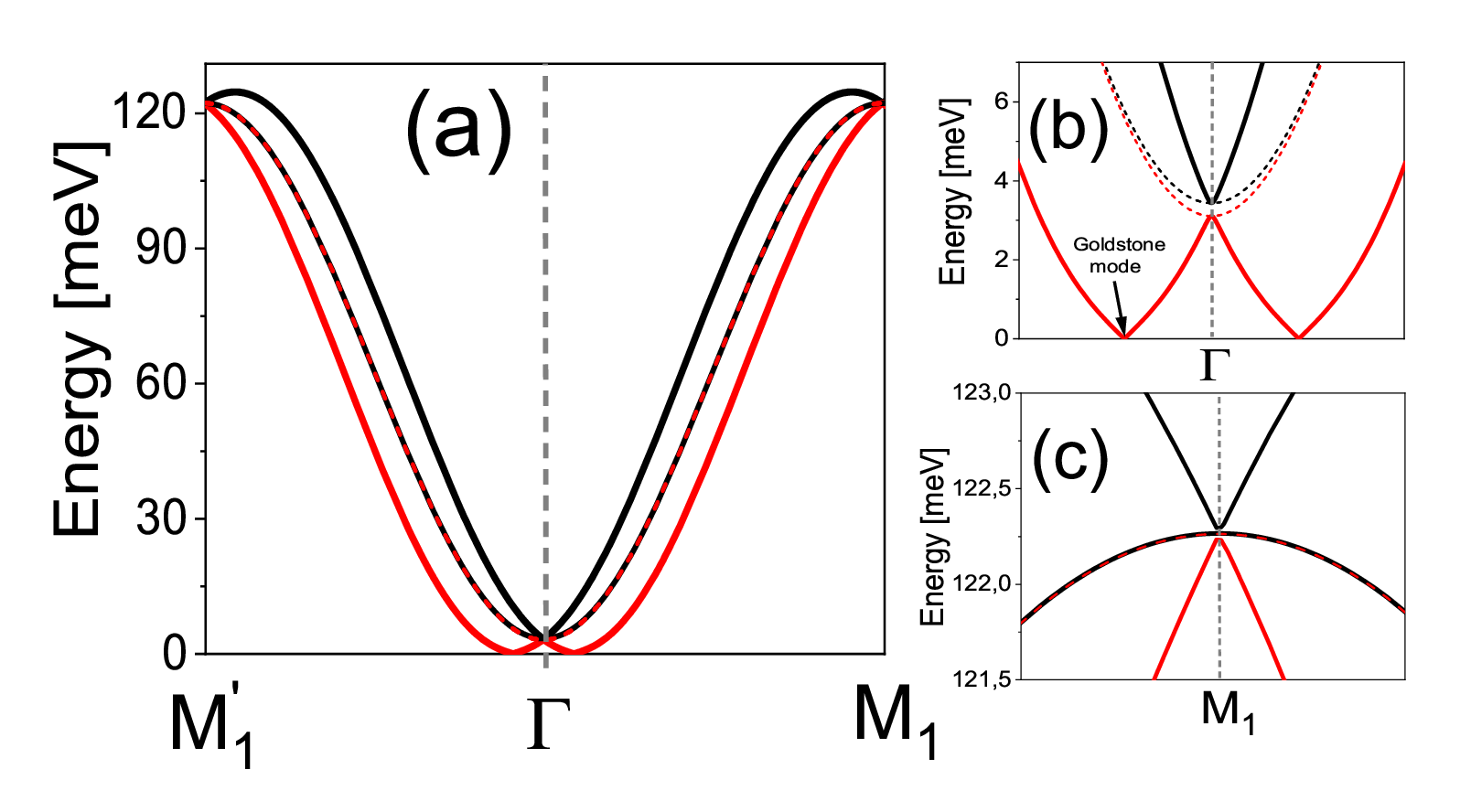}
\caption{(a) Dispersion curves of spin waves along  the  high symmetry path, M$_1\to \Gamma \to$ M$^\prime_1$ for vanishing DMI parameter, $d_{\parallel}$=0, (black and red dotted lines) and for $d_{\parallel}=5 meV$ (black and red solid lines). The zoomed in spectrum around the $\Gamma$ point is shown in (b) while that around the M$_1$ point  is shown in (c). Other parameters as described in the text.}
    \label{Fig:4}
\end{figure}

Dispersion curves of spin waves along the path with $k_x$ ranging from the point M$_1$ to the point M$^\prime _1$ via the point $\Gamma$ (center of the Brillouin zone),  M$_1\to \Gamma\to$ M$^\prime_1$, are shown in Fig.\ref{Fig:4}(a), while the spectrum in the vicinity of the  points  $\Gamma$ and M$_1$ are shown in Fig.\ref{Fig:4}(b) and Fig.\ref{Fig:4}(c), respectively. The black and red solid lines are  the dispersion curves of the two modes for nonzero DMI, while the black and red dotted lines represent the dispersion curves of the two modes for zero DMI. The latter modes are not distinguished in Fig.\ref{Fig:4}(a) and in Fig.\ref{Fig:4}(c),  where the two dotted lines overlap. However, they are resolved near the the  $\Gamma$ point, see Fig.\ref{Fig:4}(b),  where the two modes are parabolic and gaped  due to the in-plane anisotropy. In the presence of nonzero DMI, the two modes are well resolved (solid black and red lines) and the minimum of the low-energy mode is shifted out  of the $\Gamma$ point towards negative and positive $k_x$.  The spin wave energy vanishes in the minima, as indicated explicitly in Fig.\ref{Fig:4}(a), and the dispersion around these points is linear in wavevector. These are the  Goldstone modes.
This situation is similar to that in the single monolayer, where however the minimum appears only on one side, positive or negative, depending on the orientation of the spin moments. In the bilayer system there are two monolayers with opposite spin orientations, so this is why the minima occur on both sides. In turn, energy of the second mode increases with increasing $|k_x|$.
For nonzero DMI, both spin wave modes in the vicinity of the M$_1$ point display linear dispersion relation,  but with opposite group velocities. However, in the absence of DMI, the two modes are almost degenerate  and parabolic.   It is also worth to mention, that the whole spin wave spectrum in the presence of DMI  is qualitatively similar to the spectrum of 2D electronic gas with Rashba spin-orbit  coupling.
\begin{figure}[hbt!]
\centering
    \includegraphics[width=1.0\columnwidth]{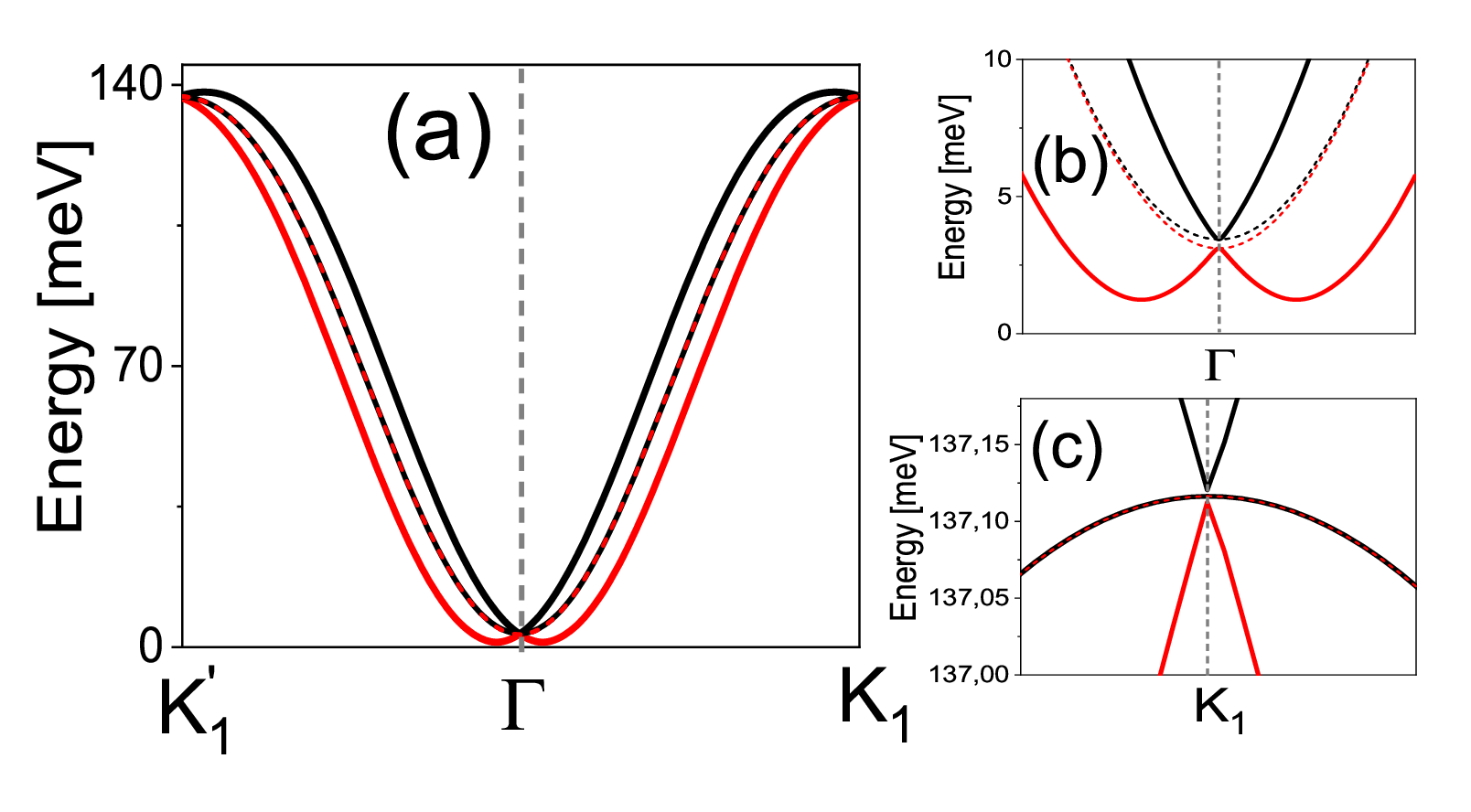}
\caption{(a) Dispersion curves of spin waves along  the  high symmetry path, K$_1\to \Gamma \to $ K$^\prime_1$ for vanishing DMI parameter, $d_{\parallel}$=0, (black and red dotted lines), and for $d_{\parallel}=5 meV$ (black and red solid lines). The zoomed in spectrum around the $\Gamma$ point is shown in (b) while that around the K$_1$ point  is shown in (c). Other parameters as described in the text.}
    \label{Fig:5}
\end{figure}

From Eqs.~(11) to (16) follows that  the dispersion relations along the $k_z$ orientation do  not depend on  the DMI parameter, so the spin wave propagation along this direction is reciprocal. In turn, the impact of DMI on the spin waves spectrum is dominant for the path M$_1\to \Gamma \to$ M$^\prime_1$, discussed above.
Between these two limiting situations, the role of DMI varies  monotonously. To show this, we present in the following the spectra along the paths K$_1\to \Gamma \to$ K$^\prime_1$, M$_2\to \Gamma \to$ M$^\prime_2$,  and along the path K$_2\to \Gamma \to$ K$^\prime_2$.   Let us consider first the path K$_1\to \Gamma \to$ K$^\prime_1$. The correponding spin wave spectrum is  shown in Fig.\ref{Fig:5}. It is clearly evident that this  spectrum is qualitatively similar to that along the path M$_1\to \Gamma \to$ M$^\prime_1$ (see Fig.\ref{Fig:4}), but with the effects due to DMI remarkably reduced in comparison to those in Fig.\ref{Fig:4}. The key difference is the absence of Goldstone modes.

\begin{figure}[hbt!]
\centering
    \includegraphics[width=1.00\columnwidth]{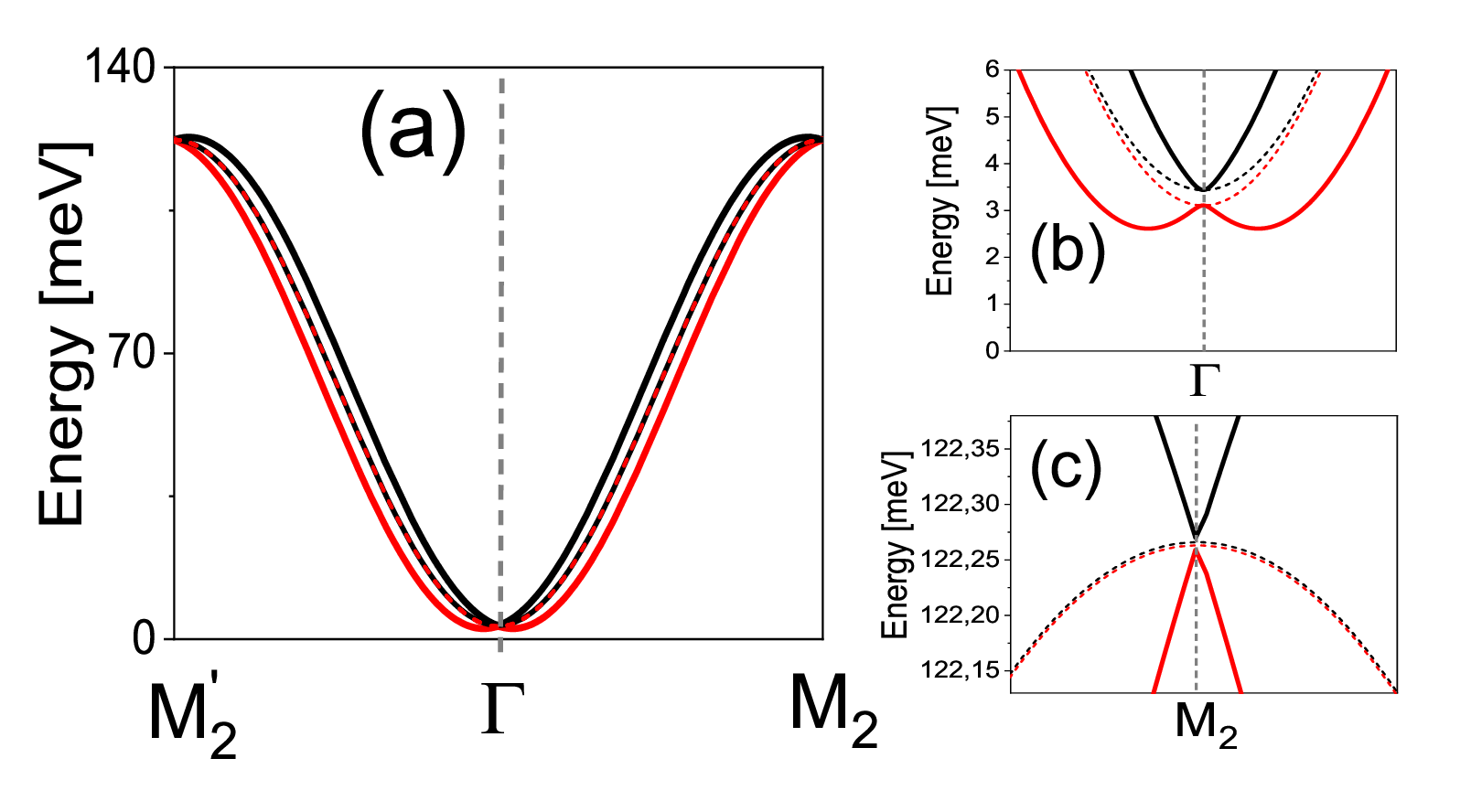}
\caption{(a) Dispersion curves of spin waves along  the  high symmetry path, M$_2\to \Gamma \to$ M$^\prime_2$ for vanishing DMI parameter, $d_{\parallel}$=0, (black and red dotted lines), and for $d_{\parallel}=5 meV$ (black and red solid lines). The zoomed in spectrum around the $\Gamma$ point is shown in (b) while that around the M point  is shown in (c). Other parameters as described in the text.}
    \label{Fig:6}
\end{figure}
\begin{figure}[h]
\centering
    \includegraphics[width=1.00\columnwidth]{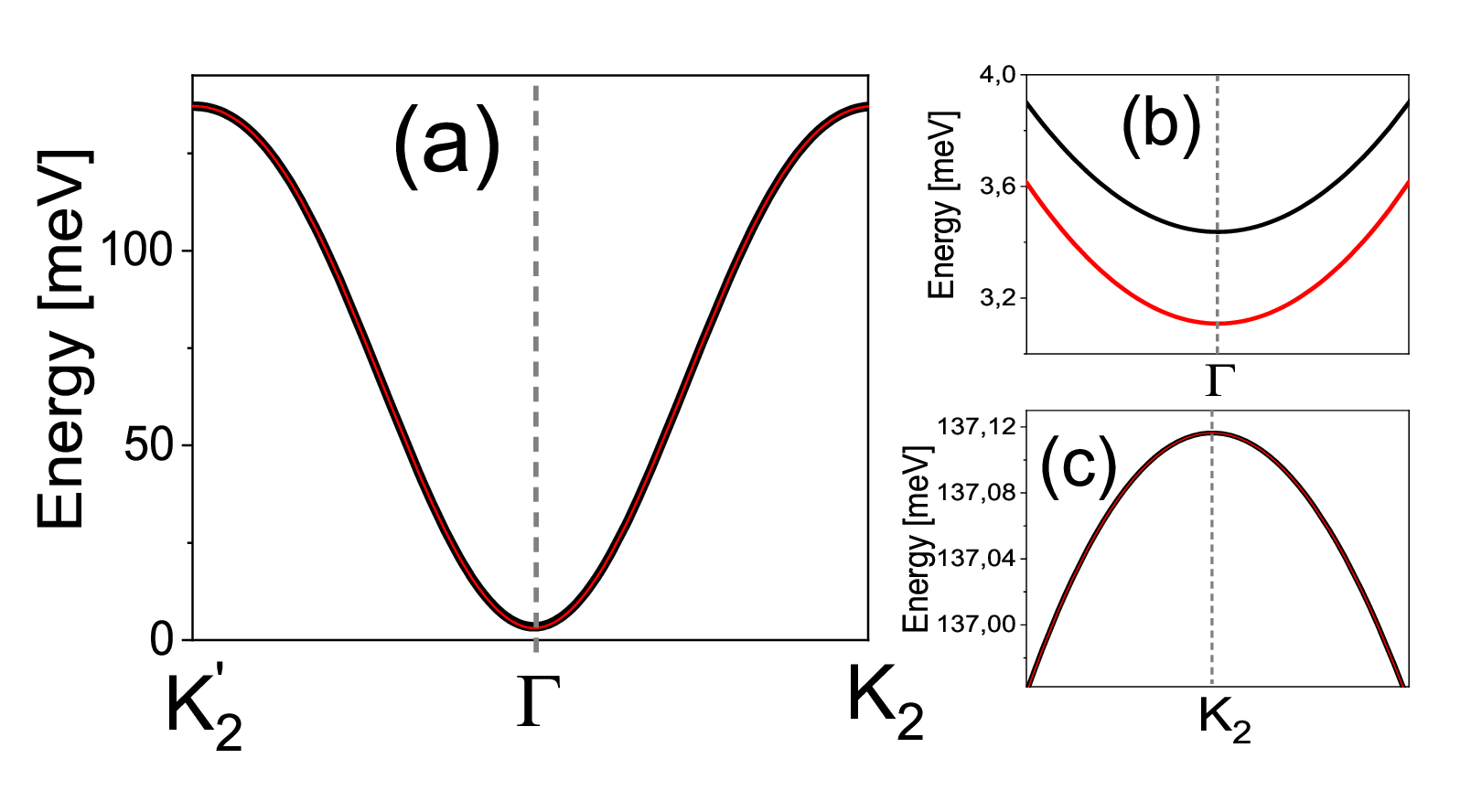}
\caption{(a) Dispersion curves of spin waves along  the  high symmetry path, K$_2\to \Gamma \to$ K$^\prime_2$ for vanishing DMI parameter, $d_{\parallel}$=0, (black and red dotted lines), and for $d_{\parallel}=5 meV$ (black and red solid lines). The zoomed in spectrum around the $\Gamma$ point is shown in (b) while that around the K point  is shown in (c). Other parameters as described in the text. Here, the spectrum is independent of DMI, and the dotted lines overlap with the corresponding solid ones.}
    \label{Fig:7}
\end{figure}

In turn, the spin wave spectra along the paths  M$_2\to \Gamma \to$ M$^\prime_2$  and  K$_2\to \Gamma \to$ K$^\prime_2$ are shown in Figs. Fig.\ref{Fig:6} and Fig.\ref{Fig:7}, respectively. In the former case, the spin wave spectrum is also qualitatively similar to that in Fig.\ref{Fig:4}, and especially to that in Fig.\ref{Fig:5}, but with the effects due to DMI reduced further (compare Fig.\ref{Fig:6} with Fig.\ref{Fig:5}). In the latter case, i.e. for the path K$_2\to \Gamma \to$ K$^\prime_2$, the spectrum is fully independent of DMI, as it is clearly visible in Fig.\ref{Fig:7}, where the two modes marked in previous figures by dotted lines overlap with the corresponding modes marked by the solid lines, so they are not resolved in Fig.\ref{Fig:7}.

\section{Summary and discussion}
We have derived analytical formulas that describe spin waves in bilayers of  TMDs  in the H and T phases, and also presented numerical results for antiferromagnetically coupled Vanadium-based dichalcogenides. The main objective was to study the influence of Dzialoshinskii-Moryia interaction  on the spin wave spectra. In these materials DMI can be induced externally by electric field due to a gate  voltage or due to substrate and/or external strains. Relatively strong DMI occurs in so-called Janus structures. The theoretical analysis is based on the spin Hamiltonian which apart from the exchange couling and DMI, also includes easy-plane and in-plane easy-axis anisotropies. The spin wave frequencies have been derived in terms of  the Hollstein-Primakoff and Bogoliubov transformations.

\section*{Acknowledgements}
This work has been supported by the Norwegian Financial Mechanism 2014- 2021 under the Polish – Norwegian Research Project NCN GRIEG “2Dtronics” no. 2019/34/H/ST3/00515.

\appendix
%%%%%%%%%%%%%%%%%%%%%%%%%%%%%%%%%%%%%%%%%%%%%%%%%%%%%%

\section{Derivation of the dispersion relation}

We consider  the antiferromagnetic  phase, which appears below the transition field to the spin-flop phase and for simplicity we assume $h=0$. In addition, we assume that the spin moments of  the bottom layer are oriented along the $z$ axis, while those of the top layer are along the $-z$ axis, see Fig.1.  Upon  the Holstein-Primakoff transformation followed by Fourier transformation, the Hamiltonian takes the form
\begin{equation}
    H=H_{\mathbf{k}}+H_{\mathbf{-k}},
\end{equation}
with
\begin{eqnarray}
    H_{\mathbf{k}} =  \sum_{\mathbf{k}}\bigg[\bigg(\frac{A_{\mathbf{k}}^+}{2}\bigg)a_{\mathbf{k},B}^+a_{\mathbf{k},B}+ \bigg(\frac{A_{\mathbf{k}}^-}{2}\bigg)a_{\mathbf{k},T}^+a_{\mathbf{k},T}
    \nonumber\\
    +B_{\mathbf{k}}a_{\mathbf{-k},T}a_{\mathbf{k},B}+ C\sum_{\alpha}a_{\mathbf{k},\alpha}a_{\mathbf{-k},\alpha}\bigg]+H.c.,
\end{eqnarray}
where  $a_{\mathbf{r},\alpha}^+$  ($a_{\mathbf{r},\alpha}$) is the bosonic  creation (anihilation) operator, while the coefficients $A_{\mathbf{k}}^{\pm}$, $B_{\mathbf{k}}$ and C  are given by the following formulas:
\begin{eqnarray}
    A_{\mathbf{k}}^{\pm}=S\bigg[2J_1\big(\gamma_\mathbf{k}-6\big)+2{\xi}J_2+\frac{D_y}{2}+D_z\nonumber\\
    \pm 4\sqrt{3}d_{||}\gamma_{\mathbf{k}}^{DM}\bigg],
    \\
     B_{\mathbf{k}}=2{\eta}^*_{\mathbf{k}}J_2S, \hspace{1cm} {\rm and} \hspace{1cm}
             C=-\frac{D_yS}{4},
\end{eqnarray}
where $+(-)$ sign in $A_{\mathbf{k}}^{\pm}$ corresponds to the bottom (top) layer.
The structure factors $\gamma_\mathbf{k}$, $\xi$,  $\gamma_{\mathbf{k}}^{DM}$ and $\eta_\mathbf{k}$ are defined  as
\begin{equation}
    \gamma_\mathbf{k}=2\bigg(\cos (k_za)+2\cos (\frac{\sqrt{3}}{2}k_xa) \cos (\frac{1}{2}k_za)\bigg),
\end{equation}
\begin{equation}
    \xi = \left\{ \begin{array}{ll}
1 & \textrm{(for T stacking)}\\
3 & \textrm{(for H stacking)},
\end{array} \right.
\end{equation}
\begin{equation}
   \gamma_{\mathbf{k}}^{DM}=\sin\bigg(\frac{\sqrt{3}}{2}k_xa\bigg)\cos\bigg(\frac{1}{2}k_za\bigg),
\end{equation}
\begin{equation}
    \eta_{\bm{k}} = \left\{ \begin{array}{ll}
1 & \textrm{(for T phase)}\\
e^{i\frac{k_xa}{\sqrt{3}}}+2e^{-i\frac{k_xa}{2\sqrt{3}}}\cos ({\frac{1}{2}k_za)} & \textrm{(for H phase).}
\end{array} \right.
\end{equation}

To diagonalize Hamiltonian Eq.(A.1), we use now Bogoliubov transfomation (for more details see Refs~\cite{owerre1,kamra,brataas})
to new bosonic operators $\Theta_{{\pm}\mathbf{k},\mu}$ and $\Theta_{{\pm}\mathbf{k},\mu}^+$, with $\mu=+,-$ indexing the two magnon modes (to be specified below). This transformation acquires the form
\begin{equation}
    %\bm{\Theta}_{\bm{\kappa}}
    \left( \begin{array}{c}
\Theta_{\mathbf{k},I\>\>\>\>\>}\\
\Theta_{\mathbf{k},II\>\>\>} \\
\Theta_{\mathbf{-k},I\>\>}^+  \\
\Theta_{\mathbf{-k},II}^+ \\
\end{array} \right)
    =\sum_{\alpha}
    \left( \begin{array}{cc}
u_{I,\alpha{\>\>\>\>\>}} & v_{I,\alpha{\>\>\>\>\>}}\\
u_{II,\alpha{\>\>\>}} & v_{II,\alpha{\>\>\>}}\\
\tilde{u}_{I,\alpha{\>\>\>\>\>}} & \tilde{v}_{I,\alpha{\>\>\>\>\>}}\\
\tilde{u}_{II,\alpha{\>\>}} & \tilde{v}_{II,\alpha{\>\>}}\\
\end{array} \right)
\left( \begin{array}{c}
a_{\mathbf{k},{\alpha\>\>\>\>}}\\
a_{\mathbf{-k},{\alpha}}^+  \\
\end{array} \right),
\end{equation}
where the Bogolubov coefficients $u_{\mu,\alpha}$ and $v_{\mu,\alpha}$ are evaluated at $\mathbf{k}$ while the coefficients $\tilde{u}_{\mu,\alpha}$ and $\tilde{v}_{\mu,\alpha}$ are evaluated at $\mathbf{-k}$,
and obey the ralations
\begin{equation}
    \sum_{\alpha}\big(|u_{\mu,\alpha}|^2+|v_{\mu,\alpha}|^2\big)=1,\>\sum_{\alpha}\big(|\tilde{u}_{\mu,\alpha}|^2+|\tilde{v}_{\mu,\alpha}|^2\big)=1.
\end{equation}

This procedure finally diagonalizes the Hamiltonian,
\begin{equation}
    H=\sum_{\mathbf{k,\mu}}{\Big(}\omega_{\mathbf{k},\mu}\Theta_{\mathbf{k},\mu}^+\Theta_{\mathbf{k},\mu}+\omega_{\mathbf{-k},\mu}\Theta_{\mathbf{-k},\mu}^+\Theta_{\mathbf{-k},\mu}\Big).
\end{equation}
Employing Eq. (A.11), one obtains  $[\Theta_{\mathbf{k},\mu},H]=\omega_{\mathbf{k},\mu}\Theta_{\mathbf{k},\mu}$, from which Eqs. (A.2) and (A.9) lead to the dispersion relation
given by Eq.(11) in the main text.

\end{document}